# THE EFFECT OF DOPANTS ON MAGNETIC PROPERTIES OF THE ORDERED $FE_{65-X}AL_{35-Y}M_{X,Y}$ ($M_{X,Y}$=GA,B,V; X,Y=5,10) ALLOYS

Voronina E. [1,5,*], Yelsukov Eu. [1,a], Korolyov A. [2,b],

Klauss H.-H. [3,c], Dellmann T. [3,d], Granovsky S. [3,4,c], Dobysheva L. [1,e]

[1] Physical-Technical Institute UrB RAS, Kirov Str. 132, 426000 Izhevsk, Russia

[2] Institute of Metal Physics UrB RAS, S. Kovalevskoi Str. 18, 620041 Ekaterinburg, Russia

[3] Institut für Festkörperphysik, TU Dresden, Helmholtzstraße 10, 01062 Dresden, Germany

[4] Lomonosov Moscow State University, Leninskie Gory 1, 119991 Moscow, Russia

[5] Institute of Physics, Kazan Federal University, Kremlyovskaya Str. 18, 420008 Kazan, Russia

[a]yelsukov@fnms.fti.udm.ru, [b]korolyov@imp.uran.ru, [c]ser@plms.phys.msu.ru, [d]dellmann@physik.tu-dresden.de, [e]lyu@otf.pti.udm.ru, *voronina@fnms.fti.udm.ru

**Abstract.** The results of X-ray diffraction, complex in-field (up to 9 T) and temperature (5–300 K) Mössbauer and magnetometric studies of the ordered $Fe_{65}Al_{35-x}M_x$ (M=Ga, B; x=0,5,10) and $Fe_{65-x}V_xAl_{35}$ (x=5,10) alloys are presented. Analysis of the magnetometry studies shows that the systems $Fe_{65}Al_{35}$ и $Fe_{65}Al_{35-x}Ga_x$ (x=5, 10) are characterized by two different magnetic states with essentially distinguishing hysteresis loops and AC susceptibility values. The temperature and external magnetic field values inducing the transition from one magnetic state to another are higher in the Ga-doped alloys than in the reference $Fe_{65}Al_{35}$ alloy. The boron addition transforms the magnetic state of the initial alloy $Fe_{65}Al_{35}$ into a ferromagnetic one exhibiting high magnetic characteristics. Substitution of V for Fe in the ternary alloys $Fe_{65-x}V_xAl_{35}$ results in reduction of magnetic characteristics and collapsing of $^{57}Fe$ hyperfine magnetic filed.

**Introduction**

Magnetic moment correlations of nanometer scale interpreted as an incommensurate spin density wave [1] and peculiar magnetotransport characteristics of the ordered $Fe_{100-x}Al_x$ alloys (25<x<35 at.%) [2] range them in a class of itinerant magnets with strong electron-electron correlations and a self-organized magnetic nanostructure. The task falls into the problem of appearance and stabilization of spiral static spin waves in itinerant magnets (in the transition metals and related alloys). *Ab*-initio theoretical studies [3,4] provide support for the spin-spiral state being the ground state at the Al concentration exceeding 30 at.%. Doping Fe-Al alloys by Ga and B will affect the change of lattice constant and V admixture will change the d-electron numbers. These ternary, Fe-Al based alloys attract interest as suitable objects to study out the origin of magnetic inhomogeneities and conditions of their formation in single-phase magnets.

**Experimental**

The materials have been produced by a special heat treatment of the originally disordered nanocrystalline alloys synthesized by mechanochemistry. Elemental powders (99.98% Fe and 99.99 % Al) and dopants have been milled in a FRITSCH P-7 planetary ball mill with vials and balls made of hardened steel (1 wt.% C and 1.5 wt.% Cr) in an argon atmosphere during 16 h. For each duration of mechanical treatment $t_{mil}$ the mass of the charged powder mixture was $m_0 = 10$ g. Under air-forced cooling the heating of the vial, balls and sample did not exceed 60 ºC. Possible ingress of the milling tools material in the sample was monitored by the measurements of the powder, vial and balls mass before and after treatment. A JAMP-10S Auger spectrometer and an MS7201M secondary-ion mass spectrometer (SIMS) were used to analyze the shape and size of the powder particles. X-ray diffraction analysis was carried out at room temperature using a DRON-3M diffractometer with monochromatic Cu $K_\alpha$ - radiation (a graphite monochromator). The phase composition was determined from X-ray diffraction patterns using Rietveld method. The $^{57}$Fe Mössbauer spectra (MS) were taken at 4.2-300 K with a conventional constant-acceleration YaGRS-4M spectrometer with a $^{57}$Co source in a Cr and Rh matrices. The magnetic measurements were carried out in the Center of magnetometry (Institute of Metal Physics, UrD RAS) by an MPMS-XL-5 (Quantum Design) SQUID magnetometer in external magnetic fields of up to 50 kOe at temperatures from 5 to 300 K. To measure the temperature dependence of susceptibility from 300 to 500 K, a home-made setup for measuring the magnetic susceptibility in small ac fields was used.

**Results and discussion**

All the samples are single-phase. Fe-Al, Fe-Al-Ga, Fe-Al-V crystallize into a B2-type superlattice, Fe-Al-B system – into a DO$_3$ one. Change in a *bcc* lattice parameter of ternary alloys (comparatively to Fe$_{65}$Al$_{35}$) is about 0.2% for the alloys admixed with 10 at.% of Ga (an increase) and V (a decrease).

| Alloy | Lattice parameter [nm] |
|---|---|
| Fe$_{65}$Al$_{35}$ | 0.2894 |
| Fe$_{65}$Al$_{30}$Ga$_5$ | 0.2897 |
| Fe$_{65}$Al$_{25}$Ga$_{10}$ | 0.2900 |
| Fe$_{65}$Al$_{30}$B$_5$ | 0.2896 |
| Fe$_{65}$Al$_{25}$B$_{10}$ | 0.2897 |
| Fe$_{60}$Al$_{35}$V$_5$ | 0.2891 |
| Fe$_{55}$Al$_{35}$V$_{10}$ | 0.2889 |

Table 1. Lattice parameters for the DO$_3$ and B2 ordered alloys

*The error is $1 \cdot 10^{-4}$ nm

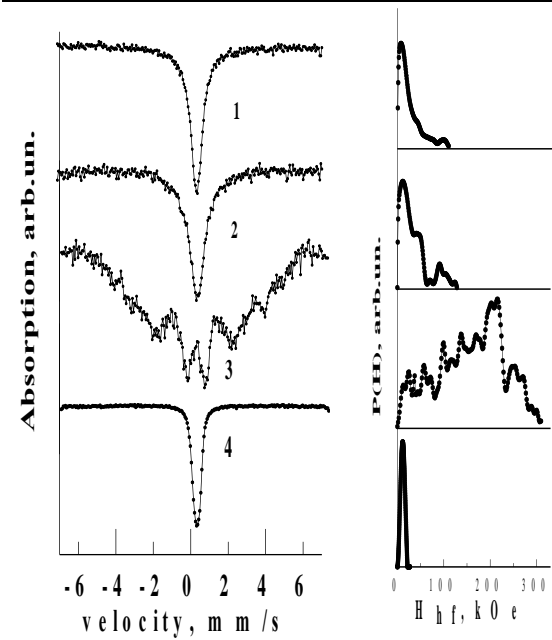

Fig.1. Mössbauer spectra taken at 80K and distributions of $^{57}$Fe hyperfine magnetic field H$_{hf}$ of the alloys: 1-Fe$_{65}$Al$_{35}$, 2- Fe$_{65}$Al$_{30}$Ga$_5$, 3- Fe$_{65}$Al$_{30}$B$_5$, 4- Fe$_{60}$Al$_{35}$V$_5$.

Mössbauer spectra and corresponding distributions of the $^{57}$Fe hyperfine magnetic field of some ordered alloys are given in Fig.1. The substitution of Ga for Al atoms in Fe$_{65}$Al$_{35}$ alloy results in a weak broadening of the singlet line (2 in Fig.1) and new components with observable magnetic splitting. The $^{57}$Fe average hyperfine magnetic field determined from the spectra taken at temperatures from 300 to 6 K changes from 0 to 145 kOe, respectively. The boron addition transforms the singlet spectrum of the Fe-Al alloy into a spectrum characteristic of a ferromagnet. Variation of Mössbauer spectra with temperature is also typical of ferromagnetic systems. Substitution of V for Fe in the ternary alloys Fe$_{65-x}$V$_x$Al$_{35}$ brings collapsing of $^{57}$Fe hyperfine magnetic filed. As it can be seen from the Fig.2, ZFC and FC loops of Fe$_{65}$Al$_{35}$ are in line, whereas the virgin magnetization curve in the hysteresis loop (see inset to Fig.2) is situated remarkably lower. Similar behaviour is observed for the alloys Fe$_{65}$Al$_{35-x}$Ga$_x$ (x=5, 10).

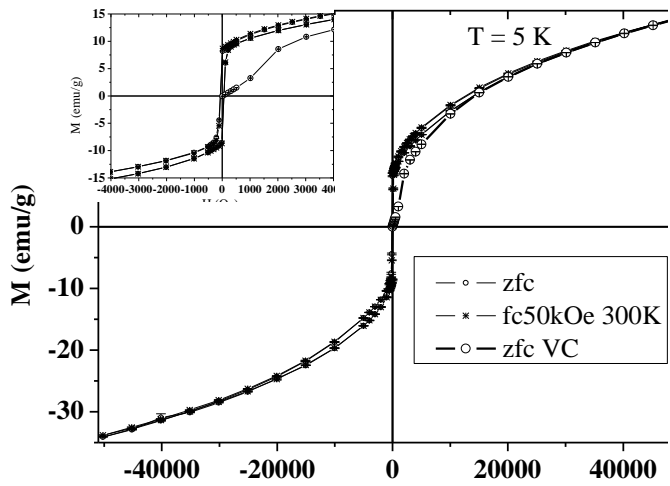

Fig.2. Magnetization curve (ZFC-VC- virgin magnetization curve) and hysteresis ZFC and FC loops of the $Fe_{65}Al_{35}$ alloy. Inset: Central part of magnetization curve and hysteresis loops

It can be supposed that after cooling under external magnetic filed $H_{ext}=0$ (ZFC regime) during magnetization of the sample the irreversible conversion of magnetic state occurs. That is, under $H_{ext} = 0$ after ZFC there is one magnetic state (1), and, after isothermal magnetization to $H_{ext} = 50$ kOe and $H_{ext}$ having been turned off, another magnetic state (2) has appeared. Following FC cycle does not give any changes in the hysteresis loop. Initial magnetic susceptibility (under $H_{ext} \sim 0$) in the magnetic state 1 is essentially lower than in the state 2.

Figure 3 presents the AC susceptibility temperature dependences. Small decrease of AC susceptibility in the first cycle is probably due to a partial transition from the higher spontaneous magnetization state 2 into the low susceptibility and magnetization state 1. However, after heating above the temperature of the transition into para-magnetic state ($T_C \cong 340$ K), a strong drop of the AC susceptibility (in comparison with the high-susceptibility state 2) is detected. In other words, heating of the sample to the temperature $T>T_C$ wipes away the high-susceptibility state 2 in-duced by external magnetic field. It can be interpreted as if magnetic moment system exists in two different magnetic states with essentially distinguishing hysteresis loops. Low-susceptibility state 1, corresponding to lower magnetization (ZFC VC), is destroyed by external magnetic field 2.5-4 T and passes to another state (stable in a broad field and temperature range) specified by a higher spontaneous magnetization. Significant increase of the AC susceptibility value in the state induced by external magnetic field indicates a substantial realignment of local magnetic moments. The temperature and external magnetic field values inducing the transition from one magnetic state to another are higher in the Ga-doped alloys than in the $Fe_{65}Al_{35}$ alloy. Considering that $T_C$ into paramagnetic state is higher than 300K ($Fe_{65}Al_{30}Ga_5$-400K, $Fe_{65}Al_{35}$-340K, Fig.3), the essential increase of the average hyperfine magnetic field in the temperature range lower than 300K points to relaxation phenomena in Mossbauer spectra and related collective magnetic moments fluctuations. It should be noted that, for the $Fe_{65}Al_{30}Ga_5$ alloy, a negligible coercive force at T=5K is observed. At the same time, assumption of superparamagnetic behaviour is not valid. Magnetization curves given as a function of H/T in Fig.4 are not in line, which should be observed for a true superparamagnetic state. ZFC and FC magnetization curves of the ordered $Fe_{65}Al_{35-x}Ga_x$ alloys (Fig.5) have a maximum as well as it was observed earlier for the ordered $Fe_{100-x}Al_x$ (x=30-35 at.%) alloys [5]. This peculiarity was explained by magnetic inhomogeneities of a nanometer size - the clusters with magnetic moments oriented parallel and antiparallel to the total magnetization. The supposition on inhomogeneities of magnetic origin is supported also by low-temperature Mössbauer studies, as far as room temperature Mössbauer spectra for the samples with $T_C$>300 K do not indicate a hyperfine magnetic splitting. It should be studied whether these inhomogeneities are conditioned by spin density waves as in the Fe-Al alloys [1] or by clusters with opposite magnetic moments.

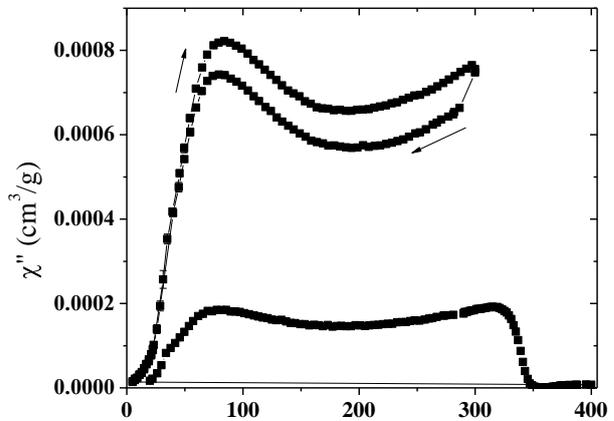

Fig. 3. Imaginary part of susceptibility $\chi''$ versus temperature of the ordered $Fe_{65}Al_{35}$ alloy. Upper curves correspond to heating to 300K followed by cooling to 5 K. Lower curve registers $\chi''$ on cooling from 400 K.

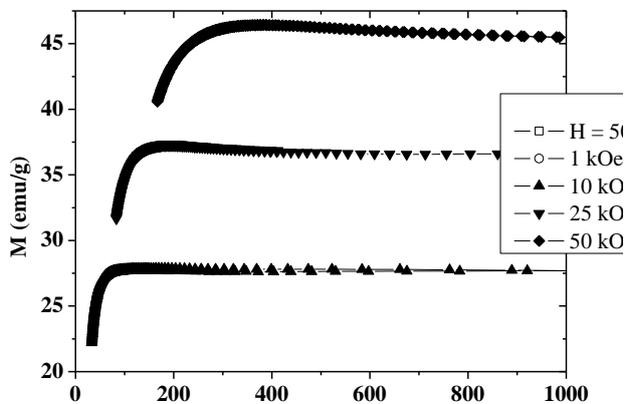

Fig.4. Magnetization curves as function of H/T for the $Fe_{65}Al_{30}Ga_5$ alloy.

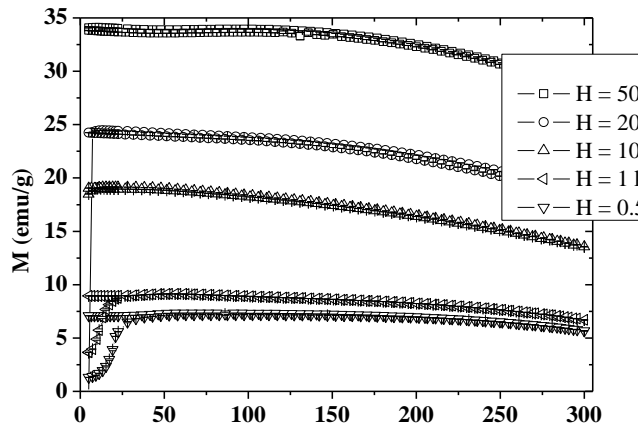

Fig.5. ZFC and FC magnetization of the ordered $Fe_{65}Al_{35}$ alloy

**Summary**

Thus, the addition of Ga to the initial alloy $Fe_{65}Al_{35}$ leads to a nonuniform magnetic state ($H_{ext}$=0 at T<$T_c$) that can be easily destroyed by external magnetic field which induces another magnetic state with a high spontaneous magnetization, stable in a broad temperature range. Boron admixture to the initial alloy $Fe_{65}Al_{35}$ results in a ferromagnetic state of the ternary alloy with high (~630K) $T_C$. Vanadium agent gives rise to abrupt diminishing of magnetic characteristics of the $Fe_{65-x}Al_{35}$ alloy, in particular, the average $^{57}Fe$ hyperfine magnetic field.

Support by RFBR (Grant № 09-02-00461) is acknowledged.